\documentclass[final,3p,times]{elsarticle}
\usepackage{amssymb}
\usepackage{multicol}
\usepackage{graphicx}
\usepackage{epsfig}
\usepackage{fancybox}
\usepackage{pstricks,pst-node,pst-text,pst-3d}
\usepackage{subfigure}
\journal{Physics Letters A}
\begin{document}
\begin{frontmatter}
\title{Stationary State Solutions of a Bond Diluted Kinetic Ising Model: An Effective-Field Theory Analysis}
\author{E. Vatansever}
\author{B.O. Aktas}
\author{Y. Yuksel}
\author{U. Akinci}
\author{H. Polat$^{\dagger}$}
\ead{hamza.polat@deu.edu.tr}
\cortext[cor1]{Corresponding Author. Tel:+90-232 4128672 Fax:+90-232 4534188}
\address{Department of Physics, Dokuz Eyl\"{u}l University, TR-35160 Izmir, Turkey}
\begin{abstract}
We have examined the stationary state solutions of a bond diluted kinetic Ising model under a time dependent oscillating magnetic field within the effective-field theory (EFT) for a honeycomb lattice $(q=3)$. Time evolution of the system has been modeled with a formalism of master equation. The effects of the bond dilution, as well as the frequency $(\omega)$ and amplitude $(h/J)$ of the external field on the dynamic phase diagrams have been discussed in detail. We have found that the system exhibits the first order phase transition with a dynamic tricritical point (DTCP) at low temperature and high amplitude regions, in contrast to the previously published results for the pure case \cite{Ling}. Bond dilution process on the kinetic Ising model gives rise to a number of interesting and unusual phenomena such as reentrant phenomena and has a tendency to destruct the first-order transitions and the DTCP. Moreover, we have investigated the variation of the bond percolation threshold as functions of the amplitude and frequency of the oscillating field.
\end{abstract}
\end{frontmatter}
\section{Introduction}\label{introduction}
Since the time in which Ising model \cite{Ising} was invented there exists a limited number of studies including the dynamic nature of the system under the influence a time dependent oscillating magnetic field. Even though the physical investigations regarding these systems bring about a lot of mathematical difficulties,
the nonequilibrium systems are in the focus of scientists because they have an unusual and interesting dynamic behavior.
By using the kinetic Ising model, the first calculation was first performed by Tom\`{e} and Oliveira \cite{tome} benefiting from Glauber-type stochastic dynamics \cite{glauber}. They investigated the time dependence of the magnetization and the temperature dependence of the dynamic order parameter (DOP) which is defined as the time averaged magnetization over a full cycle of the oscillating magnetic field within the mean field approximation (MFA). From these calculations, they obtained the dynamic phase transition (DPT) points and they contemplated the dynamic phase boundary (DPB) which separates the dynamic ordered phases from the dynamic disordered phases. They also located a dynamic tricritical point (DTCP) at which the type of phase transition changes across the DPB. Moreover, they analyzed the influence of the frequency of the external field on the DPB. Since then, a great many of theoretical effort was devoted to the DPT problem using various methods such as MFA \cite{Acharyya1, Chakrabarti, Punya}, effective-field theory (EFT) \cite{Shi, Ling, Deviren} and Monte-Carlo simulations (MC) \cite{Rao, Lo, Acharrya2, Sides, Zhu, AcharryaX}, as well as several experimental works \cite{He, Jiang, Choi, Jang, Robb}.

Besides, the lattice models including impurities have attracted considerable attention, since these models are very useful to investigate the behavior of disordered systems in nature. For this purpose, more effective models have been introduced to analyze the influence of the disorder on the thermal and magnetic properties of real  magnetic materials. Very recently, kinetic spin-1 Blume-Capel model (BC) including quenched random single-ion anisotropy has been investigated by using MFA and EFT, in order to reveal the effects of the impurities on the dynamic evolution of phase transition \cite{Yadari, Gulpinar1, Gulpinar2}. Although there exists a limited number of studies including the disorder effects under the time dependent oscillating magnetic field, the dynamic nature of the bond diluted kinetic Ising model which would exhibit an unusual dynamic behavior has not yet been investigated. Hence in this work, we intend to probe the effects of the quenched bond dilution process on the kinetic Ising model in the presence of a time-dependent  oscillating external magnetic field by using the EFT with correlations based on the exact Van der Waerden identity for a spin-1/2 system. EFT incorporates some effects of spin-spin correlations using the Van der Waerden identities and provides results that are quite superior to those obtained by MFA. The outline of the paper can be summarized as follows: The dynamic equation of motion and DOP of the bond diluted kinetic Ising model are described in the next section. The numerical results and related discussions are given in Section \ref{results} , and finally Section \ref{conclusion}  contains our conclusions.
\section{Formulation}\label{formulation}
The  kinetic Ising model is given by the time dependent Hamiltonian
\begin{equation}\label{eq1}
\mathcal{H}=-\sum_{<ij>}J_{ij}S_{i}^{z}S_{j}^{z}-\sum_{i}H(t)S_{i}^{z},
\end{equation}
where the spin variables $S_{i}^{z}=\pm1$ are defined on a honeycomb lattice $(q=3)$ and the first sum in Eq. (\ref{eq1}) is over the nearest neighbor pairs of spins. We assume that the nearest-neighbor interactions are randomly diluted on the lattice according to the probability distribution function
\begin{equation}\label{eq2}
P(J_{ij})=p\delta(J_{ij}-J)+(1-p)\delta(J_{ij}),
\end{equation}
where $p$ denotes the concentration of active bonds. The term $H(t)$ in Eq. (\ref{eq1}) is a time dependent external magnetic field which is defined as
\begin{equation}\label{eq3}
H(t)=h\cos(\omega t),
\end{equation}
where $h$ and $w=2\pi f$ represent the amplitude and the angular frequency of the oscillating field, respectively.
If the system evolves according to a Glauber-type stochastic process at a rate of $1/\tau$ which represents the transitions per unit time then the dynamic
equation of motion can be obtained as follows
\begin{equation}\label{eq4}
\tau\frac{d\langle S_{i}^{z}\rangle}{dt}=-\langle S_{i}^{z}\rangle+\left\langle\tanh\left[\frac{E_{i}+H(t)}{k_{B}T}\right]\right\rangle,
\end{equation}
where $E_{i}=\sum_{j}J_{ij}S_{j}$ is the local field acting on the lattice site $i$, and $k_{B}$ and $T$ denote the Boltzmann constant and temperature, respectively.
If we apply the differential operator technique \cite{honmura_kaneyoshi, kaneyoshi1} in Eq. (\ref{eq4}) by taking into account the random configurational averages we get
\begin{equation}\label{eq5}
\frac{dm}{dt}=-m+\left\langle\left\langle \prod_{i=1}^{q=3} \cosh(J_{ij}\nabla)+S_{i}^{z}\sinh(J_{ij}\nabla)\right\rangle\right\rangle_{r} f(x)|_{x=0},
\end{equation}
where $m=\langle\langle S_{i}^{z}\rangle\rangle_{r}$ represents the average magnetization, $\nabla=\partial/\partial x$ is a differential operator, $q$ is the coordination number of the lattice, and the inner $\langle...\rangle$ and the outer $\langle...\rangle_{r}$ brackets represent the thermal and configurational averages, respectively.
When the right-hand side of Eq. (\ref{eq5}) is expanded, the multispin correlation functions appear. The simplest approximation, and one of the most frequently adopted is to decouple these correlations according to
\begin{equation}\label{eq6}
\left\langle\left\langle
S_{i}^{z}S_{j}^{z}...S_{l}^{z}\right\rangle\right\rangle_{r}\cong\left\langle\left\langle
S_{i}^{z}\right\rangle\right\rangle_{r}\left\langle\left\langle
S_{j}^{z}\right\rangle\right\rangle_{r}...\left\langle\left\langle
S_{l}^{z}\right\rangle\right\rangle_{r},
\end{equation}
for $i\neq j \neq...\neq l$ \cite{tamura_kaneyoshi}. If we expand the right-hand side of Eq. (\ref{eq5}) with
the help of Eq. (\ref{eq6}) then we obtain the following dynamic effective-field equation of motion for the
magnetization of the bond diluted kinetic Ising model
\begin{equation}\label{eq7}
\frac{dm}{dt}=-m+\sum_{i=0}^{q=3}\lambda_{i}m^{i},
\end{equation}
where the coefficients $\lambda_{i}$  can be easily calculated by employing the mathematical relation $\exp(\alpha \nabla)f(x)=f(x+\alpha)$. Eq. (\ref{eq7})
describes the nonequilibrium behavior of the system in the EFT formalism. Moreover, the time dependence of magnetization can be one of two types according to whether they obey the following property or not
\begin{equation}\label{eq8}
m(t)=-m(t+\pi/\omega).
\end{equation}
A solution satisfying Eq. (\ref{eq8}) is called symmetric solution and it corresponds to a paramagnetic (P) or disordered phase. In this type solution, the time
dependent magnetization oscillates around zero value. The second type of solutions which does not satisfy  Eq. (\ref{eq8}) is called nonsymmetric solution which corresponds to ferromagnetic (F)  (i.e. ordered) phase  where the time dependent magnetization oscillates around nonzero value. We can also mention that the differential equation derived in Eq. (\ref{eq7}) is a type of initial value problem and this equation extends to $m^3$ for kinetic Ising model. Each term in this differential equation gives contribution to the solution because the value of $m$ which is calculated at each time step is related to the previous $m$ value.

The time averaged magnetization over a full cycle of the oscillating magnetic field acts as DOP which is defined as follows \cite{tome}
\begin{equation}\label{eq9}
Q=\frac{\omega}{2\pi}\oint m(t)dt,
\end{equation}
where $m(t)$ is a stable and periodic function. The behavior of DOP as a function of the temperature for selected Hamiltonian parameters is obtained
by solving Eqs. (\ref{eq7}) and (\ref{eq9}) numerically. We solve these equations by combining the fourth order Runge-Kutta method (RK4) with the Trapeze integration method. After some transient steps which depends on the value of the set of Hamiltonian parameters, the system is expected to be in a stationary state where the magnetization satisfies the condition $m(t)=m(t+2\pi/\omega)$ which is necessary to calculate the DOP.

\section{Results and Discussion}\label{results}
In this section, we will discuss the dynamic nature of the critical phenomena in a bond diluted kinetic Ising model and we will also touch upon the mechanism behind the DPT. It is well known that one cannot write the free energy expression including the presence of a time dependent oscillating magnetic field. Hence, we  cannot use the free energy expression to determine the type of the phase transition. In order to overcome this problem, we will follow a procedure described briefly below. Since the time average of the magnetization over a full cycle of the external magnetic field acts as the DOP then the variation of it as a function of the temperature is checked, in order to determine the type of phase transition. If the DOP decreases continuously  to zero, this DPT point is classified as of second order. On the other hand, if it vanishes discontinuously with a finite jump, this transition is classified as of first order. From this point of view and after detailed analysis to find DPB that separates the dynamic ordered phases from the dynamic disordered phases, we have found that the behavior of the system changes dramatically with varying Hamiltonian parameters. In order to investigate the effects of the Hamiltonian parameters on the dynamic behavior of the system, we plot the phase diagrams in various planes.

\begin{figure}
\center
\includegraphics[width=12cm]{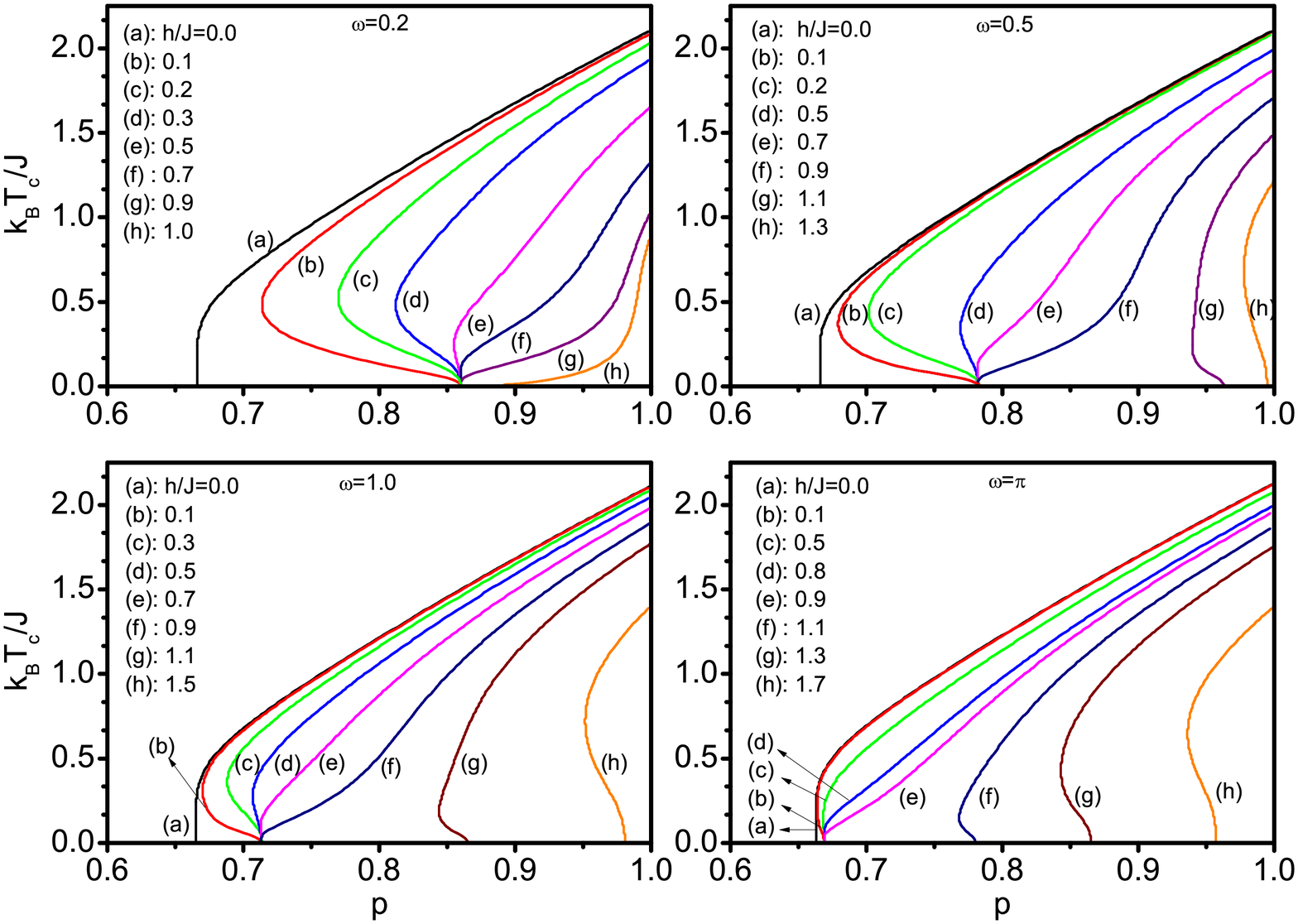}\\
\caption{The phase diagrams of the kinetic bond diluted system in  $(k_{B}T_{c}/J-p)$ plane with some selected values of frequency of the oscillating magnetic field (a) $w=0.2$, (b) $w=0.5$, (c) $w=1.0$, and (d) $w=\pi$. The letters accompanying each curve represent the amplitude $h/J$ of the external field.}\label{fig1}
\end{figure}

In Fig. \ref{fig1}, we plot the phase diagrams in $(k_{B}T_{c}/J-p)$ plane for $\omega=0.2, 0.5, 1.0$ and $\pi$ with some selected values of $h/J$. As seen in Fig. (\ref{fig1}), as $p$ value decreases then the F phase region gets narrower. As the concentration of active bonds is decreased then the energy contribution which comes from the spin-spin interactions gets smaller.  Hence, the system can undergo a DPT at lower critical temperatures, due to the energy originating from the temperature and (or) magnetic field overcomes ferromagnetic the spin-spin interactions. Therefore, the F regions in  $(k_{B}T_{c}/J-p)$ plane get narrower.  In addition, increasing $h/J$ values reduce the dynamic critical temperature whereas as $\omega$ increases then  the F phase region gets wider in the $(k_{B}T_{c}/J-p)$ plane. These observations can also be explained by the well known mechanisms underlying the DPT phenomena. In other words, increasing field amplitude facilitates the phase transition due to the increasing energy coming from the oscillating magnetic field which tends to align the spins in its direction whereas an increasing field frequency gives rise to a growing phase delay between the magnetization and magnetic field (i.e. the magnetization cannot follow the oscillating magnetic field) and this makes the occurrence of the DPT difficult. Hence, as a result of this mechanism, critical temperature increases as $\omega$ increases. We can mention that depending on the applied field and frequency, the system may exhibit an interesting behavior. Namely, the system exhibits a reentrant behavior for $w=0.2$ and for the low amplitudes, such as $h/J\leq0.5$ while the reentrant region in $(k_{B}T_{c}/J-p)$ plane extends to higher amplitudes as $w$ increases. Moreover, it is also observed that the reentrant behavior originates at higher concentrations with increasing frequency $\omega$. As seen from the curve with $h/J=0.1, \omega=1.0$ and $p=0.70$,  corresponding to the lower left panel Fig. (1) as the temperature increases starting from a value corresponding to P phase the system undergoes a second order phase transition  due to the thermal agitations and remains at a F state for a while. As the temperature  increases further then the system undergoes another second order phase transition between F and P phases.

\begin{figure}
\center
\includegraphics[width=12cm]{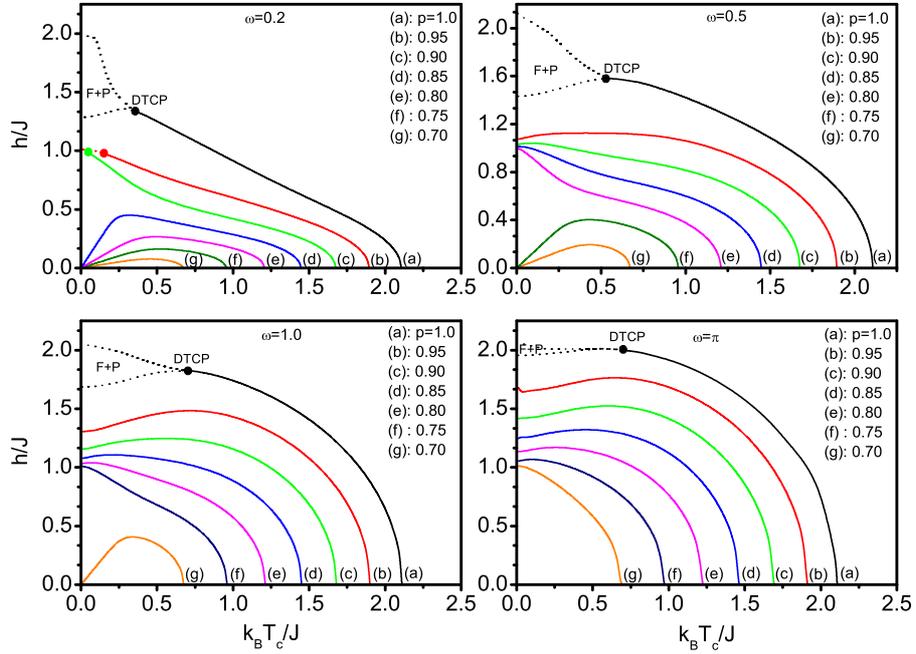}\\
\caption{The phase diagrams of the kinetic bond diluted system in  $(k_{B}T_{c}/J-h/J)$ plane with some selected values of frequency of the oscillating magnetic field (a) $w=0.2$, $w=0.5$, $w=1.0$, and $w=\pi$. Solid and dotted lines represents the second and the first order phase transitions, respectively, and the solid circles represent the $DTCP$s. The letters on the curves denote the concentration $p$ of active bonds.}\label{fig2}
\end{figure}

The physical background of the critical phenomena observed in the bond diluted kinetic Ising model depends on the amplitude of external field, as well as the dilution process. Based on this fact, we illustrate the effect of $h/J$  on the phase diagrams for some selected Hamiltonian parameters in Fig. \ref{fig2}. Before investigating the influence of the bond dilution process on the thermal and magnetic properties of the kinetic Ising model we shall note that the pure kinetic Ising model exhibits meta-stable behavior, in contrast to the previously published results \cite{Ling}. For example, the pure system exhibits the first order phase transitions and also a coexistence region at low temperature and high amplitude values where the F and P phases overlap. In addition, there exists a DTCP on the DPB at which the second order and first order phase transition lines are separated from each other. Although the behavior of the pure kinetic Ising model has been studied by Ling et al.\cite{Ling}, they didn't report any evidence regarding the first order phase transition and DTCP, as well as any coexistence region  since they didn't analyze the global phase diagrams in temperature versus amplitude space in detail. Besides, one can see from Fig. \ref{fig2} that the system exhibits a reentrant behavior of second order where the two successive second order phase transitions take place in the low amplitude and low temperature regions for low concentrations (i.e. the curves labeled "(g)" in Fig. \ref{fig2}) which disappears as $\omega$ increases. If the concentration value increases then the behavior of the system begins to resemble the pure kinetic Ising model and a first order reentrant behavior with a meta-stable (F+P) phase appears where the critical properties of the system depends on the initial value of the magnetization. The reentrant behavior disappears with increasing frequency at the low temperature and low amplitude regions and the region of the ordered F phase widens. This is an expected result, since increasing the field frequency causes a growing phase delay between the magnetization and magnetic field and this makes the occurrence of the DPT difficult, as a result of this mechanism the DPB gets wider. It is clear that the results presented in Fig. \ref{fig2} are completely consistent with the results shown in Fig. \ref{fig1}. Physical mechanism mentioned above can be briefly explained as follows: If one keeps the system  in one well of a Landau type double well potential, a certain amount of energy originating from magnetic field is necessary to achieve a dynamic symmetry breaking.  If the amplitude of the applied field is less than the required amount then the system oscillates in one well. In this situation, the magnetization does not change its sign. In other words, the system oscillates around a nonzero value. This region is dynamically ordered phase. When the temperature increases, the height of the barrier between the two wells decreases. As a result of this, the less amount of magnetic field is necessary to push the system from one well to another and hence the magnetization can change its sign for this amount of field. Consequently, the time averaged magnetization over a full cycle of the oscillating field becomes zero $(Q=0)$. In addition, we can mention that for a bond diluted kinetic Ising model, increasing the effect of the bond disorder in the system by diluting the active bond concentration,  takes the ground state energy of system far from that of the pure system as a result of decreasing amount of ferromagnetic spin-spin interactions which makes the transition from one well to another easy.

\begin{figure}
\center
\includegraphics[width=12cm]{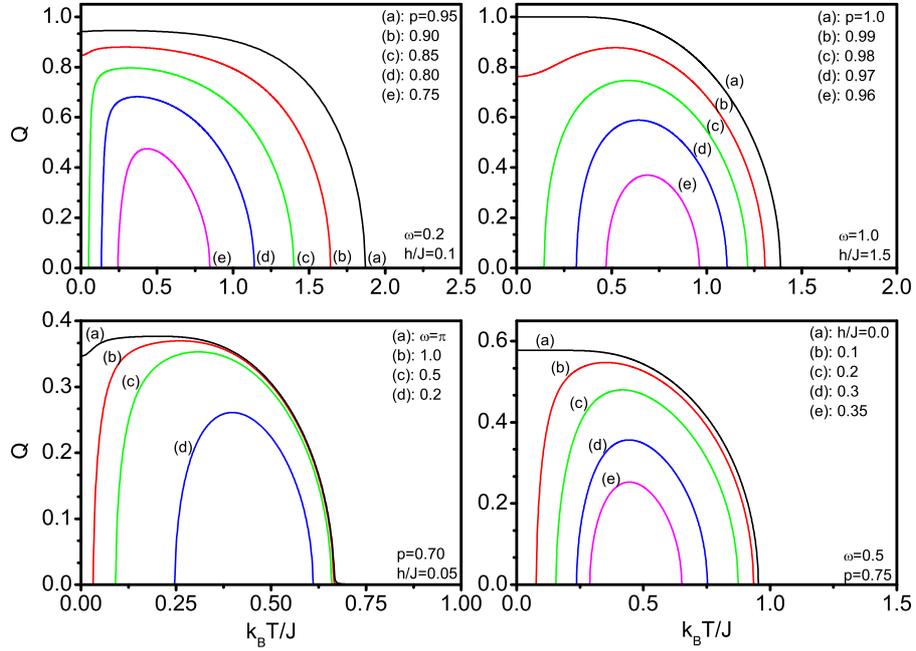}\\
\caption{Thermal variations of the $DOP$ curves as functions of the selected Hamiltonian parameters $h/J$, $w$ and $p$ where the letters on the curves denote the value of the corresponding Hamiltonian parameter.}\label{fig3}
\end{figure}

In Fig.\ref{fig3}, we plot the thermal variations of the DOP curves corresponding to the phase diagrams depicted in Figs. \ref{fig2} and \ref{fig3}. As seen in the upper left and right panels in Fig. \ref{fig3}, as the concentration decreases then the system tends to exhibit reentrant behavior of second order. For example, as the temperature increases starting from zero then a second order phase transition from a disordered phase to an ordered phase at low temperature regions is followed by another phase transition of the second order which takes place from an ordered phase to a disordered phase at higher temperature regions for decreasing $p$ values. This situation holds for both weak and strong frequency (or amplitude) values which can be clearly seen from the upper left and right panels in Fig. \ref{fig3}. Moreover, as $p$ decreases then the ferromagnetic region gets narrower and after a certain concentration value $(p^{*})$  which is called as the bond concentration threshold, the system cannot exhibit an ordered phase anymore. Furthermore, the dependencies of the DOP versus temperature curves as a function of the oscillating field frequency are represented in the lower left panel in Fig. \ref{fig3} for $p=0.7$ and $h/J=0.05$. As shown in this figure, value of the DOP increases when the frequency approaches the static case in the limit $(\omega\rightarrow 0)$.  In order to clarify the effect of the amplitude of the external magnetic field on the dynamic order parameter of the system, we represent the variation of the DOP with temperature in the lower right panel in Fig. \ref{fig3} for $\omega=0.5$ and $p=0.75$, corresponding  to the low amplitude regions of the upper right panel in Fig. \ref{fig2}. As seen in this figure, as we increase the value of the field amplitude $h/J$ then the system tends to exhibit a second order reentrance.

\begin{figure}
\center
\includegraphics[width=10cm]{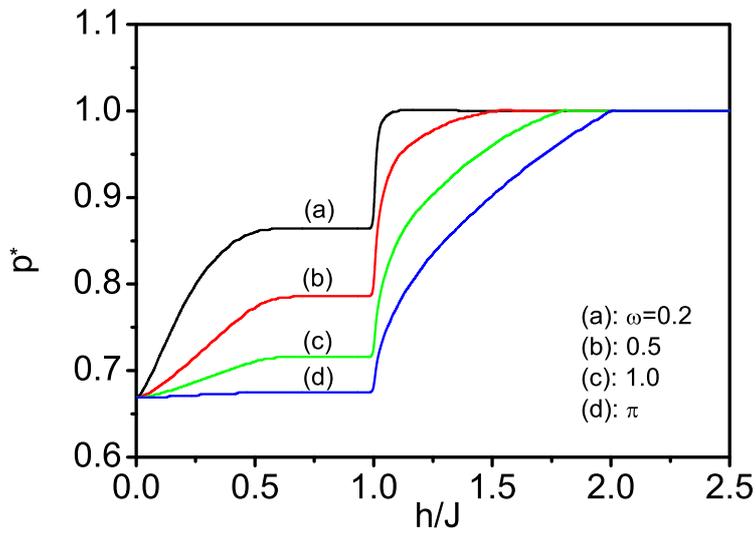}\\
\caption{Variation of the bond percolation threshold $p^{*}$ as a function of the oscillating field amplitude $h/J$ for some selected values of the oscillation frequency $\omega$.  The letters on each curve denote the frequency value of the external field.}\label{fig4}
\end{figure}

As a final investigation, let us examine the variation of the bond percolation threshold $p^{*}$ as a function of $h/J$ with some selected values of $\omega$ which is depicted in Fig. \ref{fig4}. As expected, for $h/J=0$ we get the value of bond percolation threshold for the static system \cite{Kaneyoshi2} which is independent of $\omega$. In addition, at relatively small oscillation frequency values such as $\omega=0.2, 0.5$ and $1.0$, value of $p^{*}$ increases gradually then exhibits a plateau which gets wider as $\omega$ increases. For $h/J>1.0$, $p^{*}$ value increases rapidly and saturates at $p^{*}=1.0$. Furthermore, for $h/J>1.0$, as $\omega$ increases we get $p^{*}=1.0$ for higher amplitude values. Hence we should note that the bond percolation threshold value $p^{*}$ strictly depends on a kind of competition effect which originates from the collaboration of the nearest neighbor interactions with the oscillation frequency against the amplitude of the external field.

\section{Conclusion}\label{conclusion}
In conclusion, we have investigated the stationary states of the bond diluted kinetic Ising model in the presence of the oscillating magnetic magnetic field within the framework of EFT on a two dimensional honeycomb lattice. For this purpose, the Glauber stochastic dynamics has been used to describe the time evolution of the system. We have explored the global phase diagrams of the system, including the reentrant phase transitions, as well as DTCP's and after some detailed analysis we have found that the system exhibits reentrant phenomena for some certain values of amplitude and frequency of the external field. We have also found that F phase regions get expanded with decreasing amplitude which is more evident at low frequencies. Moreover, we can say that decreasing concentration of bonds causes the destruction of the first order phase transitions and coexistence regions.  After a certain value of concentration, the first order phase transitions turn into the second order phase transitions and consequently, DTCP's disappear for all frequency values.

EFT method takes the standard mean field predictions one step forward by taking into account the self spin correlations which means that the thermal fluctuations are partially considered within the framework of EFT. Although all of the observations reported in this work shows that EFT can be successfully applied to such nonequilibrium systems in the presence of quenched bond disorder, the true nature of the physical facts underlying the observations displayed in the present work (especially the origin of the coexistence phase) may be further understood with an improved version of the
present EFT formalism which can be achieved by attempting to consider the multi spin correlations which originate when expanding the spin identities. We believe that this attempt could provide a treatment beyond the present approximation.

As a conclusion, we hope that the results obtained in this work would shed light on the further investigations of the dynamic nature of the critical phenomena in disordered systems and would be beneficial from both theoretical and experimental points of view.

\section*{Acknowledgments}
The numerical calculations reported in this paper were performed at T\"{U}B\.{I}TAK ULAKB\.{I}M (Turkish agency), High Performance and Grid Computing Center (TRUBA Resources) and this study has
been completed at Dokuz Eylul University, Graduate School of Natural and Applied Sciences. One of the authors (B.O.A.) would like to thank the Turkish Educational
Foundation (TEV) for partial financial support.

\newpage

\newpage

\newpage

\newpage

\end{document}